\documentclass[conference,a4paper]{IEEEtran}
\IEEEoverridecommandlockouts
\usepackage{cite}
\usepackage{amsmath,amssymb,amsfonts}
\usepackage{algorithmic}
\usepackage{graphicx}
\usepackage{textcomp}
\usepackage{xcolor}
\usepackage{tikz}
\usepackage{pgfplots}
\usepackage{booktabs,array,multirow}
\definecolor{uni_apfelgruen}{cmyk}{.5, 0, 1, 0}
\definecolor{uni_mittelblau}{cmyk}{1, 0.4, 0, 0}
\definecolor{uni_gelb}{cmyk}{0, 0.1, 1, 0}
\definecolor{uni_rot}{cmyk}{0, 1, 1, 0}
\definecolor{uniblauHell}{RGB}{0,190,255}
\definecolor{uniblauDunkel}{RGB}{0,65,145}
\definecolor{unigrau}{RGB}{51,51,51}
\definecolor{darkgray176}{RGB}{176,176,176}
\definecolor{lavenderplot}{RGB}{191,148,228}
\definecolor{coralplot}{RGB}{255,127,80}
\definecolor{cyanplot}{RGB}{37,219,168}
\usepgfplotslibrary{groupplots}
\usetikzlibrary{matrix,arrows.meta}

\newcommand{\PreserveBackslash}[1]{\let\temp=\\#1\let\\=\temp}
\newcolumntype{C}[1]{>{\PreserveBackslash\centering}p{#1}}
\newcolumntype{R}[1]{>{\PreserveBackslash\raggedleft}p{#1}}
\newcolumntype{L}[1]{>{\PreserveBackslash\raggedright}p{#1}}

\def\BibTeX{{\rm B\kern-.05em{\sc i\kern-.025em b}\kern-.08em
    T\kern-.1667em\lower.7ex\hbox{E}\kern-.125emX}}
\begin{document}
\title{Learning to exploit z-Spatial Diversity for Coherent Nonlinear Optical
Fiber Communication\\
}

\author{\IEEEauthorblockN{Sebastian Jung, Tim Uhlemann, Alexander Span, Maximilian Bauhofer and Stephan ten Brink}
\IEEEauthorblockA{\textit{Institute of Telecommunications}\\
\textit{University of Stuttgart}\\
\textit{Pfaffenwaldring 47, 70569 Stuttgart, Germany}\\
\textit{\{jung,uhlemann,span,bauhofer,tenbrink\}@inue.uni-stuttgart.de}}
}
\maketitle

\begin{abstract}
Higher-order solitons inherently possess a spatial periodicity along the propagation axis.
The pulse expands and compresses in both, frequency and time domain.
This property is exploited for a bandwidth-limited receiver by sampling the optical signal at two different distances.
Numerical simulations show that when pure solions are transmitted and the second (i.e., further propagated) signal is also processed, a significant gain in terms of required receiver bandwidth is obtained.
Since all pulses propagating in a nonlinear optical fiber exhibit solitonic behavior given sufficient input power and propagation distance, the above concept can also be applied to spectrally efficient Nyquist pulse shaping and higher symbol rates.
Transmitter and receiver are trainable structures as part of an autoencoder, aiming to learn a suitable predistortion and post-equalization using both signals to increase the spectral efficiency.
\end{abstract}


\section{Introduction}

The ever-increasing demand for data means that the resources available in today's communications systems must either be used more efficiently or expanded.
Optical fibers in general provide very high bandwidths. For instance, the C-band, commonly used for communication, covers a total bandwidth of around \(4.4~\mathrm{THz}\).
However, the exploitation of this spectrum, or, at least the expansion to higher bandwidths is limited by the required signal processing, e.g., the bandwidths of digital-to-analog (DAC) and analog-to-digital (ADC) conversion, respectively.
These components typically support bandwidths of $100\,\mathrm{GHz}$ to $200\,\mathrm{GHz}$ maximum, which is only a small fraction of the available channel spectrum.
Furthermore, as the optical fiber is a nonlinear medium, the signal broadens as it propagates, occupying additional spectral resources. Summing up, bandwidth limitation is the main driver for the performance degradation, or more positively, one of the highest potentials for higher-power optical communications systems.

One approach to face this bottleneck, is to split the C-band into multiple channels, i.e., in a wavelength-division multiplexing (WDM) setup.
This way, the limited bandwidth of DACs and ADCs is sufficient to cover the entire C-band.
Typically, around $100$ different channels per polarization \cite{Dar2016} are used in parallel. 
However, there is still a bandwidth limitation, or rather a difference between channel spacing and bandwidth, to reduce the interference of neighboring channels.

In this work, the idea is to exploit additional and available bandwidth of the optical fiber, while still relying on further limited converters. The approach is to take advantage of the propagation properties of signals on the optical fiber, i.e., the solitonic propagation of any waveform given a sufficiently high launch power and distance. Note that the two concepts do not exclude one another.

\begin{figure}
    \centering
    \begin{tikzpicture}[rotate=90,scale=1.2]

\definecolor{darkgray176}{RGB}{176,176,176}
\pgfplotsset{compat=newest}
\begin{axis}[
tick align=outside,
tick pos=left,
xtick pos=top,
xlabel near ticks,
x grid style={darkgray176},
xlabel={frequency \(\displaystyle f-f_0\) in \(\displaystyle \mathrm{GHz}\)},
xmin=-15000000000, xmax=15000000000,
xtick style={color=black},
xtick={-15000000000,-10000000000,-5000000000,0,5000000000,10000000000,15000000000},
xticklabels={\ensuremath{-}1.5,\ensuremath{-}1.0,\ensuremath{-}0.5,0.0,0.5,1.0,1.5},
y dir=reverse,
y grid style={darkgray176},
ylabel={Propagation distance \(\displaystyle z\) in \(\displaystyle \mathrm{km}\)},
ymin=0, ymax=1000,
ytick style={color=black},
width=.5\linewidth,
height=.8\linewidth,
xtick={-15e9,-10e9,-5e9,0,5e9,10e9,15e9},
xticklabels={\(-15\),,,\(0\),,,\(15\)},
scaled x ticks=false,
yticklabel style={rotate=-90,font=\scriptsize},
xticklabel style={rotate=-90,font=\scriptsize},
ylabel style={rotate=180, font=\scriptsize},
ytick={0},
yticklabels={\(0\hspace*{-.4em}\rightarrow\)},
ylabel shift=-12pt,
xlabel style={font=\scriptsize}
]
\addplot graphics [includegraphics cmd=\pgfimage,xmin=-15000000000, xmax=15000000000, ymin=1000, ymax=0] {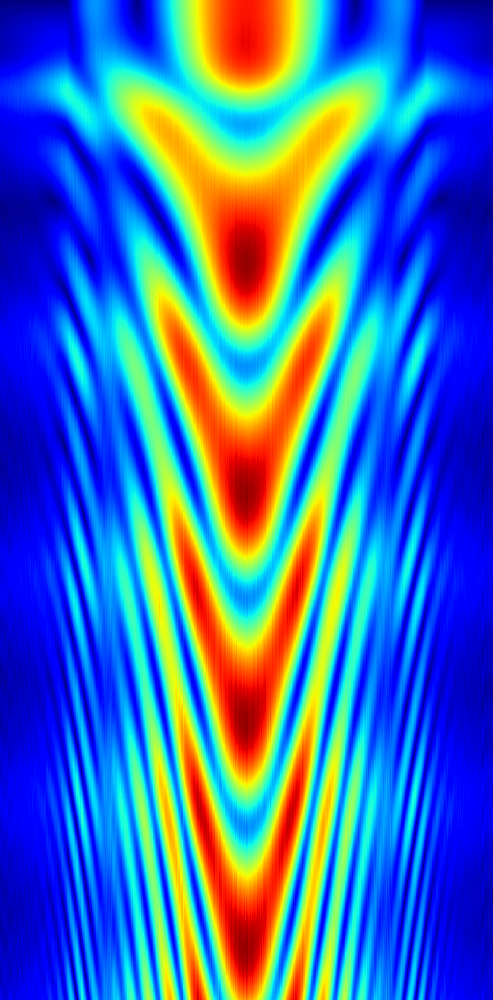};
\end{axis}
\draw[thick, black] (1.94,.3) -- (3.15,.3);
\draw[thick,black] (3.15,.1) rectangle (3.55,.5);
\draw[thick,black] (3.15,.5) -- (3.55,.1);
\draw (3.45,.4) node{\tiny D};
\draw (3.25,.2) node{\tiny A};
\draw[thick,-{Triangle[length=1mm, width=.66mm]}] (3.55,.3) -- (3.7,.3);
\draw (3.65,.3) node[anchor=south]{\scriptsize\(\mathbf{y}_2\)};
\draw (3.02,.3) node[anchor=east]{\scriptsize\(\ell_2\)};
\draw[very thick, black] (1.44,.3) ellipse[x radius=.5,y radius=.2];

\draw[thick, black] (1.94,.85) -- (3.15,.85);
\draw[thick,black] (3.15,1.05) rectangle (3.55,.65);
\draw[thick,black] (3.15,1.05) -- (3.55,.65);
\draw (3.45,.95) node{\tiny D};
\draw (3.25,.75) node{\tiny A};
\draw[thick,-{Triangle[length=1mm, width=.66mm]}] (3.55,.85) -- (3.7,.85);
\draw (3.65,.85) node[anchor=south]{\scriptsize\(\mathbf{y}_1\)};
\draw (3.02,.85) node[anchor=east]{\scriptsize\(\ell_1\)};
\draw[very thick, black] (1.44,.85) ellipse[x radius=.5,y radius=.2];
\end{tikzpicture}
    \caption{Intensity of higher-order solitons: Motivation for using \(z\)-spatial diversity}
    \label{fig::solitons}
\end{figure}
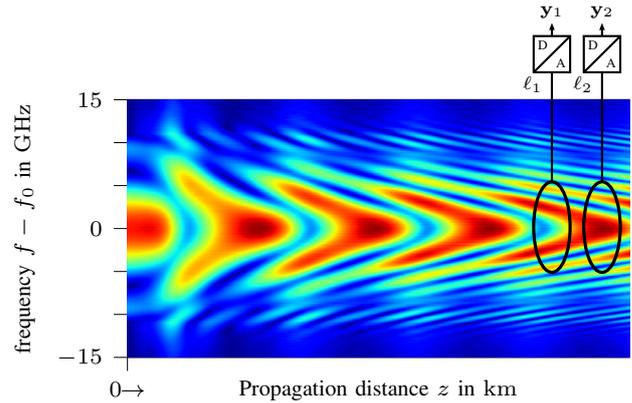

Higher-order solitons posses a spatial periodicity in bandwidth that can be exploited by sampling the signal at two different points in propagation distance, i.e., the $z$-axis. Thus, information that may be outside the converters' bandwidth at one point may have re-entered the available spectrum at another point in $z$-direction.
This principle is shown in Fig. \ref{fig::solitons}.
Since there is no known analytical solution to combine the two $z$-spatially different signals obtained in this way, we apply and optimize a neural network (NN) to perform this task.
More precisely, a trainable Autoencoder \cite{Oshea2017} is used to (i) pre-distort the single transmit signal and (ii) combine and post-equalize the received signals.

It is already well known that Digital-Backpropagation (DBP) is able to compensate for all channel effects if no bandwidth limitation is present and signal-noise interaction can be neglected \cite{Pare1996, Essiambre2005}.
Therefore, in this work, we use DBP as a baseline algorithm that does not have the ability to exploit $z$-spatial diversity.
In a previous work, we have already shown that the performance degradation due to the bandwidth limitation of DBP can be partially circumvented by an appropriate Autoencoder-based pre-processing of the transmit signal \cite{Uhlemann2022}.
Therefore, we use the results of that work to design an Autoencoder that not only compensates for, but also exploits the spectral propagation characteristics of optical signals.
By having designed and trained the spoken Autoencoder, we achieved a gain of over \(+0.4\,\frac{\mathrm{bit}}{\mathrm{s}\cdot\mathrm{Hz}}\) in terms of spectral efficiency.

The remainder of the paper is organized as follows.
Sec. \ref{sec::basics} provides the used channel model and performance metric as well as the basic idea of the concept.
The whole communication setup is provided in Sec. \ref{sec::setup} and the obtained results are presented in Sec. \ref{sec::results}.
Finally, in Sec. \ref{sec::summary} a summary of the work and an outlook are given.

\section{Basics\label{sec::basics}}
Propagation in a single mode fiber is governed by the Nonlinear Schroedinger Equation (NLSE).
Since we want to  exploit the  solitonic  properties  of  propagating  impulses, we assume ideal distributed Raman amplification (IDRA).
Therefore, the attenuation term is dropped and the amplified spontaneous emission (ASE) noise is added instead.
\begin{align}
    \frac{\partial q\left(t,z\right)}{\partial z} = \mathrm{j}\frac{\beta_2}{2}\frac{\partial^2 q\left(t,z\right)}{\partial t^2}-\mathrm{j}\gamma\left\vert q\left(t,z\right)\right\vert^2 q\left(t,z\right) + n\left(t,z\right)
\end{align}
Here \(q\left(t,z\right)\) is the optical baseband signal with time \(t\) and distance \(z\) and \(\beta_2\) and \(\gamma\) are the dispersion and nonlinearity parameters, respectively.
This differential equation is typically solved numerically using the Split-Step Fourier Method (SSFM).
The used implementation can be found in \cite{sionna}.
A standard-single mode fiber (S-SMF) is used as the communication medium in the C-band around \(f_\text{0}=193.55\,\mathrm{THz}\).

The power of the noise term \(n\left(t,z\right)\) is calculated from \cite{Muga2009}
\begin{align}
    \sigma_{\mathrm{n},\mathrm{ase}}^2\left(\Delta z\right) = n_\text{sp}\left(f_0\right)hf_0\alpha\Delta zB
\end{align}
with spontaneous emission factor \(n_\text{sp}\), Planck's constant \(h\), attenuation \(\alpha\) and bandwidth \(B\).

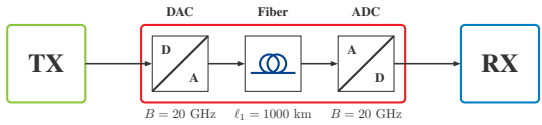
\begin{figure}
    \centering
    \resizebox{0.8\linewidth}{!}{\begin{tikzpicture}
	\draw[color=uni_apfelgruen, rounded corners=3, ultra thick] (-1.8,-.3) rectangle (-3.9,1.8);
	\draw[color=unigrau] (-2.85, .75) node{\huge \textbf{TX}};
	\draw[color=uni_mittelblau, rounded corners=3, ultra thick] (8.3,-.3) rectangle (10.4,1.8);
	\draw[color=unigrau] (9.35, .75) node{\huge \textbf{RX}};
	\draw[color=uni_rot,rounded corners=3,ultra thick] (-.3,-.3) rectangle (6.8,1.8);
	\foreach \i in {0,1,2} {
		\draw[color=unigrau, thick,fill=white] (2.5*\i,0) rectangle (2.5*\i+1.5,1.5);
	}

	\foreach \i in {0,1}{
		\draw[very thick,color=unigrau,-{Triangle[length=2mm,width=1.33mm]}] (1.5+2.5*\i,.75) -- (1.5+2.5*\i+1,.75);
	}

	\draw[ultra thick,color=unigrau] (0,0) -- (1.5,1.5);
	\draw[color=unigrau] (0.375,1.125) node{\textbf{D}};
	\draw[color=unigrau] (1.125,.375) node{\textbf{A}};
    \draw[color=unigrau] (.75,1.8) node[anchor=south]{\textbf{DAC}\strut};
    \draw[color=unigrau] (.75,-.3) node[anchor=north]{\(B=20~\mathrm{GHz}\)\strut};
	
	\draw[color=uniblauDunkel,ultra thick] (2.65,.5) -- (3.85,.5);
	\draw[color=uniblauDunkel, ultra thick] (3.25-.1,.75) circle (.25);
	\draw[color=uniblauDunkel, ultra thick] (3.25+.1,.75) circle (.25);
    \draw[color=unigrau] (3.25,1.8) node[anchor=south]{\textbf{Fiber}\strut};
    \draw[color=unigrau] (3.25,-.3) node[anchor=north]{\(\ell_1=1000~\mathrm{km}\)\strut};

	\draw[ultra thick,color=unigrau] (5,0) -- (6.5,1.5);
	\draw[color=unigrau] (5.375,1.125) node{\textbf{A}};
	\draw[color=unigrau] (6.125,.375) node{\textbf{D}};
    \draw[color=unigrau] (5.75,1.8) node[anchor=south]{\textbf{ADC}\strut};
    \draw[color=unigrau] (5.75,-.3) node[anchor=north]{\(B=20~\mathrm{GHz}\)\strut};
	
	\draw[very thick,color=unigrau,-{Triangle[length=2mm,width=1.33mm]}] (-1.8,.75) -- (0,.75);
	\draw[very thick,color=unigrau,-{Triangle[length=2mm,width=1.33mm]}] (6.5,.75) -- (8.3,.75);
\end{tikzpicture}}
    \caption{System model}
    \label{fig:channel_model}
\end{figure}

The transmitters and receivers used in this work are bandwidth-limited due to the DAC and ADC.
Fig.~\ref{fig:channel_model} shows the channel model including the bandwidth limitation.
Since the Kerr-effect  causes  spectral  broadening, information is lost at the receiver.

This bandwidth limitation is also the reason why conventional systems degrade in performance when the Kerr-effect becomes the dominant degradation, i.e., at higher input power.

In our setup, the DAC and ADC have a passband-bandwidth of \(B=20\,\mathrm{GHz}\) and are assumed to be ideal filters with a brickwall spectrum.

The simulations are performed at a frequency of \(f_\text{sim} = 1\,\mathrm{THz}\) with a symbol rate of \(R_\mathrm{S}= 20\,\mathrm{GBd}\).
This results in a resolution of \(f_\text{sim}/R_\mathrm{S}=50\) samples per symbol.

\subsection{Performance metric}
In contrast to our previous work, we use soft decision spectral efficiency \(\eta\), which is an achievable information rate obtained by evaluating the output of a Gaussian mixture model (GMM)-based demapper with its corresponding likelihoods.
This metric is upper bounded by Shannon's limit \cite{Kramer2015}.

\subsection{Spatial diversity}
$z$-spatial diversity describes the approach to use the spatial periodicity of higher order solitons to circumvent the bandwidth limitation of DAC and ADC.
Hereby, the signal is sampled at two different distances along the propagation axis $z$.

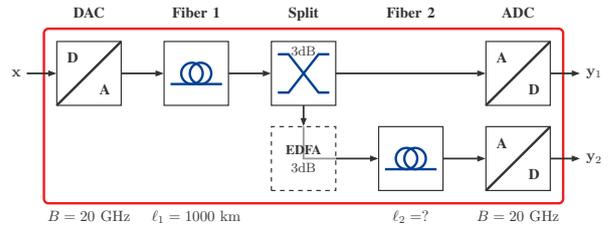
\begin{figure}
    \centering
    \resizebox{.9\linewidth}{!}{\begin{tikzpicture}
	\draw[color=uni_rot,rounded corners=3, ultra thick] (-.3,-2.3) rectangle (11.8,1.8);
	\foreach \i in {0,1,2,4} {
		\draw[color=unigrau, thick,fill=white] (2.5*\i,0) rectangle (2.5*\i+1.5,1.5);
	}

	\foreach \i in {3,4}{
		\draw[color=unigrau,thick,fill=white] (2.5*\i,-2) rectangle (2.5*\i+1.5,-.5);
	}

	\foreach \i in {0,1}{
		\draw[very thick,color=unigrau,-{Triangle[length=2mm,width=1.33mm]}] (1.5+2.5*\i,.75) -- (1.5+2.5*\i+1,.75);
	}
	\draw[very thick,color=unigrau,-{Triangle[length=2mm,width=1.33mm]}] (1.5+5,.75) -- (1.5+7.5+1,.75);
	\foreach \i in {2,3}{
		\draw[very thick,color=unigrau,-{Triangle[length=2mm,width=1.33mm]}] (1.5+2.5*\i,-1.25) -- (1.5+2.5*\i+1,-1.25);
	}

	\draw[very thick, color=unigrau] (5.75,-.5) -- (5.75,-1.25) -- (6.5,-1.25);
	\fill[color=white,opacity=.5] (5,-2) rectangle (6.5,-.5);
	\draw[color=unigrau,thick,dashed] (5,-2) rectangle (6.5,-.5);
	\draw[color=unigrau] (5.75,-1.05) node{\small\textbf{EDFA}};
	\draw[color=unigrau] (5.75,-1.45) node{\small\(3\mathrm{dB}\)};

	\draw[ultra thick,color=unigrau] (0,0) -- (1.5,1.5);
	\draw[color=unigrau] (0.375,1.125) node{\textbf{D}};
	\draw[color=unigrau] (1.125,.375) node{\textbf{A}};
    \draw[color=unigrau] (.75,1.8) node[anchor=south]{\textbf{DAC}\strut};
    \draw[color=unigrau] (.75,-2.3) node[anchor=north]{\(B=20~\mathrm{GHz}\)\strut};
	
	\draw[color=uniblauDunkel,ultra thick] (2.65,.5) -- (3.85,.5);
	\draw[color=uniblauDunkel, ultra thick] (3.25-.1,.75) circle (.25);
	\draw[color=uniblauDunkel, ultra thick] (3.25+.1,.75) circle (.25);
    \draw[color=unigrau] (3.25,1.8) node[anchor=south]{\textbf{Fiber 1}\strut};
    \draw[color=unigrau] (3.25,-2.3) node[anchor=north]{\(\ell_1=1000~\mathrm{km}\)\strut};

	\draw[color=uniblauDunkel, ultra thick] (5.15,.3) -- (5.4,.3) -- (6.1,1.2) -- (6.35,1.2);
	\draw[color=uniblauDunkel, ultra thick] (5.15,1.2) -- (5.4,1.2) -- (6.1,.3) -- (6.35,.3);
	\draw[color=unigrau] (5.75,1.27) node{\small\(3\mathrm{dB}\)};

	\draw[color=unigrau,very thick,-{Triangle[length=2mm,width=1.33mm]}] (5.75,0) -- (5.75,-.5);
    \draw[color=unigrau] (5.75,1.8) node[anchor=south]{\textbf{Split}\strut};

	\draw[color=uniblauDunkel,ultra thick] (7.65,-1.5) -- (8.85,-1.5);
	\draw[color=uniblauDunkel, ultra thick] (8.25-.1,-1.25) circle (.25);
	\draw[color=uniblauDunkel, ultra thick] (8.25+.1,-1.25) circle (.25);
    \draw[color=unigrau] (8.25,1.8) node[anchor=south]{\textbf{Fiber 2}\strut};
    \draw[color=unigrau] (8.25,-2.3) node[anchor=north]{\(\ell_2=?\)\strut};

	\draw[ultra thick,color=unigrau] (10,0) -- (11.5,1.5);
	\draw[color=unigrau] (7.875+2.5,1.125) node{\textbf{A}};
	\draw[color=unigrau] (8.625+2.5,.375) node{\textbf{D}};
	
	\draw[ultra thick,color=unigrau] (10,-2) -- (11.5,-.5);
	\draw[color=unigrau] (7.875+2.5,1.125-2) node{\textbf{A}};
	\draw[color=unigrau] (8.625+2.5,.375-2) node{\textbf{D}};
    \draw[color=unigrau] (10.75,1.8) node[anchor=south]{\textbf{ADC}\strut};
    \draw[color=unigrau] (10.75,-2.3) node[anchor=north]{\(B=20~\mathrm{GHz}\)\strut};

	\draw[very thick,color=unigrau,-{Triangle[length=2mm,width=1.33mm]}] (-.7,.75) node[anchor=east]{\(\mathbf{x}\)} -- (0,.75);
	\draw[very thick,color=unigrau,-{Triangle[length=2mm,width=1.33mm]}] (11.5,.75) -- (12.2,.75) node[anchor=west]{\(\mathbf{y}_1\)};
	\draw[very thick,color=unigrau,-{Triangle[length=2mm,width=1.33mm]}] (11.5,-1.25) -- (12.2,-1.25) node[anchor=west]{\(\mathbf{y}_2\)};
\end{tikzpicture}}
    \caption{Realization of spatial diversity}
    \label{fig::sd_model}
\end{figure}

The signal is sent over an optical fiber that spans the communication distance of \(\ell_1=1000\,\mathrm{km}\).
Afterwards, a \(3\,\mathrm{dB}\)-coupler splits the signal.
One part is passed directly to the digital signal processing (DSP) after the ADC, the other part is optionally amplified using an Erbium-doped fiber amplifier (EDFA), to compensate for the splitter loss, and sent through a second fiber of length \(\ell_2\), ADC, and finally passed to the DSP.
The entire structure is shown in Fig.~\ref{fig::sd_model}.
Two hyperparameters remain to be optimized: the fiber length \(\ell_2\) and the use of the EDFA.
These will be optimized later on.

\subsection{Spatial diversity with solitons}
A proof-of-concept of the $z$-spatial diversity setup was done.
There, second-order solitons were sent through the spatial diversity setup at $5\,\mathrm{dBm}$, with the eigenvalues $\left\{\lambda_1=0.5\mathrm{j}, \lambda_2=\mathrm{j}\right\}$ and a periodic length of $z_\mathrm{p}=483\,\mathrm{km}$.
The first sampling of the signal is performed at \(\ell_1=1000\,\mathrm{km}\).
Compensation of the splitter loss using an EDFA was mandatory since solitons were sent.
In addition, ideal Raman amplification is assumed.

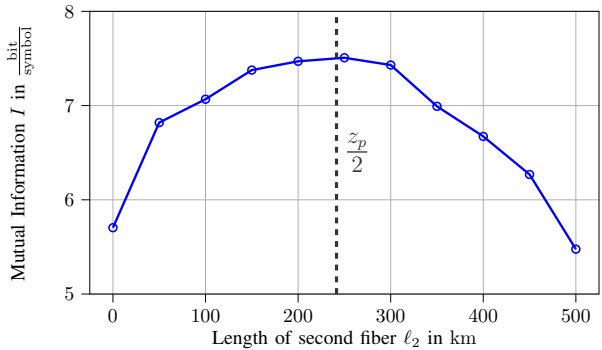
\begin{figure}
    \centering
    \resizebox{.9\linewidth}{!}{\begin{tikzpicture}
\tikzset{
	invisible/.style={opacity=0},
	visible on/.style={alt={#1{}{invisible}}},
	alt/.code args={<#1>#2#3}{%
		\alt<#1>{\pgfkeysalso{#2}}{\pgfkeysalso{#3}} %
	},
}
\begin{axis}[
legend cell align={left},
legend style={
	fill opacity=0.8,
	draw opacity=1,
	text opacity=1,
	at={(0.815,0.65)},
	anchor=south,
	draw=black,
	nodes={scale=0.6, transform shape}
},
tick align=outside,
tick pos=left,
x grid style={white!69.0196078431373!black},
xlabel={Length of second fiber $\ell_{2}$ in \(\mathrm{km}\)},
xmajorgrids,
xmin=-24.9895000002347, xmax=524.999500000011,
xtick style={color=black},
y grid style={white!69.0196078431373!black},
ylabel={Mutual Information \(I\) in  \(\frac{\mathrm{bit}}{\mathrm{symbol}}\)},
ymajorgrids,
ymin=5, ymax=8,
ytick style={color=black},
width=1.2\linewidth,
height=.75\linewidth
]

\draw[dashed, unigrau, ultra thick] (axis cs: 241.5,0) -- (axis cs: 241.5,8);
\draw[dashed, unigrau, ultra thick] (axis cs: 241.5,6.5) node[anchor=west]{\LARGE\(\frac{z_p}{2}\)};

\addplot [very thick, blue, mark=o, mark options={solid, thick}]
table {%
0.00999999046325684 5.70413970947266
50 6.81940889358521
100 7.06747531890869
150 7.3765721321106
200 7.46978139877319
250 7.50725507736206
300 7.43023681640625
350 6.9920392036438
400 6.67251825332642
450 6.26862812042236
500 5.47617673873901
};

\end{axis}

\end{tikzpicture}}
    \caption{Simulation results using pure solitons}
    \label{fig::soliton_gain}
\end{figure}

Simulation results for combining both paths using a neural network (NN) are shown in Fig. \ref{fig::soliton_gain}.
There, the mutual information versus the length of the second fiber \(\ell_2\) is plotted.
A clear optimum is visible.
It occurs at half of the periodic length, i.e., at the length where the impulse is at the other extreme of its periodic cycle.

\section{Setup\label{sec::setup}}
\begin{figure}
    \centering
    \resizebox{\linewidth}{!}{\begin{tikzpicture}
	
	\draw[color=white,ultra thin] (-1.8,1.805) -- (14.3,1.805);
	\draw[color=uni_apfelgruen,rounded corners=3,ultra thick] (-.3,-.3) rectangle (12.8,1.8);
	
	\foreach \i in {0,1,2,3,4} {
		\draw[thick,color=unigrau,fill=white] (2.75*\i,0) rectangle (2.75*\i+1.5,1.5);
	}

	\foreach \i in {0,1,2,3}{
		\draw[very thick,color=unigrau,-{Triangle[length=2mm,width=1.33mm]}] (1.5+2.75*\i,.75) -- (1.5+2.75*\i+1.25,.75);
	}
	\draw[very thick,color=unigrau,-{Triangle[length=2mm,width=1.33mm]}] (12.5,.75) -- (13.2,.75) node[anchor=west]{\(\mathbf{x}\)\strut};

	\draw[ultra thick, color=unigrau] (.15,.75) --(1.35,.75);
	\draw[ultra thick, color=unigrau] (.75,.15) --(.75,1.35);
	\fill[color=uniblauDunkel] (1.06,1.06) circle (.07);
	\fill[color=uniblauDunkel] (1.06,.44) circle (.07);
	\fill[color=uniblauDunkel] (.44,1.06) circle (.07);
	\fill[color=uniblauDunkel] (.44,.44) circle (.07);
	\draw[color=unigrau] (.75,1.8) node[anchor=south]{\textbf{Mapper}\strut};
    \draw[color=unigrau] (.75,-.3) node[anchor=north]{\(M=256~\text{symbols}\)\strut};
	
	\draw[thick,unigrau,fill=uni_gelb!30!white] (2.75,0) rectangle (4.25,1.5);
	\draw[color=unigrau] (3.5,.75) node{\large\textbf{NN}};
	\draw[color=unigrau] (3.5,1.8) node[anchor=south]{\textbf{Predistortion}\strut};
    \draw[color=unigrau] (3.5,-.3) node[anchor=north]{architecture \cite{Freire2021}\strut};
	
	\draw[ultra thick,color=unigrau,-{Triangle[length=3mm,width=2mm]}] (5.75,.15) -- (5.75,1.35);
	\draw[color=unigrau] (5.75,.75) node[anchor=west]{1:50};
	\draw[color=unigrau] (6.25,1.8) node[anchor=south]{\textbf{Upsampler}\strut};
    \draw[color=unigrau] (6.25,-.3) node[anchor=north]{\(R_\mathrm{S}\rightarrow f_\text{sim}\)\strut};
	
	\begin{axis}[width=88.5,height=88.5,ticks=none,xmin=-2.5,xmax=2.5,at={(2700,0)},axis line style={draw=none}]
		\addplot[mark=none,domain=-2.3:2.3,samples=1000,color=uniblauDunkel,ultra thick] {0.564*e^(-x^2)};
	\end{axis}
	\draw[color=unigrau] (9,1.8) node[anchor=south]{\textbf{Pulse Shaper}\strut};
    \draw[color=unigrau] (9,-.3) node[anchor=north]{RRC, roll-off: \(0\)\strut};
	
	\draw[color=unigrau] (11.75,.75) node{\(\frac{\mathbf{x}\cdot\sqrt{P_\mathrm{L}}}{\Vert \mathbf{x}\Vert}\)};
	\draw[color=unigrau] (11.75,1.8) node[anchor=south]{\textbf{Normalizer}\strut};
    \draw[color=unigrau] (11.75,-.3) node[anchor=north]{\(-5~\mathrm{dBm}\dots5\mathrm{dBm}\)\strut};
	
	\draw[color=unigrau] (2.125,.75) node[anchor=south]{\(\mathbf{s}\)\strut};
	\draw[color=unigrau] (4.875,.75) node[anchor=south]{\(\tilde{\mathbf{s}}\)\strut};
	\draw[color=unigrau] (7.625,.75) node[anchor=south]{\(\mathbf{x}_\mathrm{US}\)\strut};
	\draw[color=unigrau] (10.375,.75) node[anchor=south]{\(\mathbf{x}_\mathrm{PS}\)\strut};
	
\end{tikzpicture}}\\
    \vspace*{.3cm}
    \resizebox{\linewidth}{!}{\begin{tikzpicture}
	
\draw[color=white,ultra thin] (-1.8,2.805) -- (14.3,2.805);
	\draw[color=uni_mittelblau,rounded corners=3, ultra thick] (-.3,-1.3) rectangle (12.8,2.8);
	
	\foreach \i in {3,4} {
		\draw[thick,color=unigrau,fill=white] (2.75*\i,0) rectangle (2.75*\i+1.5,1.5);
	}

	\foreach \i in {0,1,2}{
		\draw[thick,color=unigrau,fill=white] (2.75*\i,1) rectangle (2.75*\i+1.5,2.5);
		\draw[thick,color=unigrau,fill=white] (2.75*\i,.5) rectangle (2.75*\i+1.5,-1);
	}

	\foreach \i in {0,1}{
		\draw[very thick,color=unigrau,-{Triangle[length=2mm,width=1.5mm]}] (1.5+2.75*\i,1.75) -- (1.5+2.75*\i+1.25,1.75);
		\draw[very thick,color=unigrau,-{Triangle[length=2mm,width=1.5mm]}] (1.5+2.75*\i,-.25) -- (1.5+2.75*\i+1.25,-.25);
	}
	\draw[very thick,color=unigrau,-{Triangle[length=2mm,width=1.5mm]}] (7,1.75) -- (8.25,.75);
	\draw[very thick,color=unigrau,-{Triangle[length=2mm,width=1.5mm]}] (7,-.25) -- (8.25,.75);
	\draw[very thick,color=unigrau,-{Triangle[length=2mm,width=1.5mm]}] (9.75,.75) -- (11,.75);
	\draw[very thick,color=unigrau,-{Triangle[length=2mm,width=1.5mm]}] (-.7,1.75) node[anchor=east]{\(\mathbf{y}_1\)\strut} -- (0,1.75);
	\draw[very thick,color=unigrau,-{Triangle[length=2mm,width=1.5mm]}] (-.7,-.25) node[anchor=east]{\(\mathbf{y}_2\)\strut} -- (0,-.25);

	\draw[color=unigrau] (.75,1.75) node{\textbf{CDC}};
	\draw[color=unigrau] (.75,-.25) node{\textbf{CDC}};
	\draw[color=unigrau] (.75,2.8) node[anchor=south]{\textbf{Compensation}\strut};
    \draw[color=unigrau] (.75,-1.3) node[anchor=north]{ideal\strut};
	
	\draw[scale =0.1, domain = -1.85:1.85, variable = \x, uniblauDunkel, ultra thick]  plot ({\x*3 +15+20},{-\x*\x*\x*\x + 23});
    \draw[scale =0.1, domain = -1.85:1.85, variable = \x, uniblauDunkel, ultra thick]  plot ({\x*3 +15+20},{-\x*\x*\x*\x+3});
	\draw[color=unigrau] (3.5,2.8) node[anchor=south]{\textbf{Lowpass}\strut};
    \draw[color=unigrau] (3.5,-1.3) node[anchor=north]{RRC, roll-off: \(0\)\strut};
 
	\draw[ultra thick,color=unigrau,-{Triangle[length=3mm,width=2mm]}] (5.75,2.35) -- (5.75,1.15);
	\draw[color=unigrau] (5.75, 1.75) node[anchor=west]{50:1};
	\draw[ultra thick,color=unigrau,-{Triangle[length=3mm,width=2mm]}] (5.75,.35) -- (5.75,-.85);
	\draw[color=unigrau] (5.75, -.25) node[anchor=west]{50:1};
	\draw[color=unigrau] (6.25,2.8) node[anchor=south]{\textbf{Downsampler}\strut};
    \draw[color=unigrau] (6.25,-1.3) node[anchor=north]{\(f_\text{sim}\rightarrow R_\mathrm{S}\)\strut};

	\draw[thick,color=unigrau,fill=uni_gelb!30!white] (8.25,0) rectangle (9.75,1.5);
	\draw[color=unigrau] (9,.75) node{\large\textbf{NN}};
	\draw[color=unigrau] (9,2.8) node[anchor=south]{\textbf{Combining}\strut};
    \draw[color=unigrau] (9,-1.3) node[anchor=north]{architecture \cite{Freire2021}\strut};
	
	\draw[thick,color=unigrau,fill=uni_gelb!30!white] (11,0) rectangle (12.5,1.5);
	\draw[ultra thick, color=unigrau] (11.15,.75) --(12.35,.75);
	\draw[ultra thick, color=unigrau] (11.75,.15) --(11.75,1.35);
	\draw[color=uniblauDunkel,ultra thick] (12.06+0.06,1.06+0.06) -- (12.06-0.06,1.06-0.06);
	\draw[color=uniblauDunkel,ultra thick] (12.06+0.06,1.06-0.06) -- (12.06-0.06,1.06+0.06);
	\draw[color=uniblauDunkel,ultra thick] (12.06+0.06,.44+0.06) -- (12.06-0.06,.44-0.06);
	\draw[color=uniblauDunkel,ultra thick] (12.06-0.06,.44+0.06) -- (12.06+0.06,.44-0.06);
	\draw[color=uniblauDunkel,ultra thick] (11.44+0.06,1.06+0.06) -- (11.44-0.06,1.06-0.06);
	\draw[color=uniblauDunkel,ultra thick] (11.44-0.06,1.06+0.06) -- (11.44+0.06,1.06-0.06);
	\draw[color=uniblauDunkel,ultra thick] (11.44+0.06,.44+0.06) -- (11.44-0.06,.44-0.06);
	\draw[color=uniblauDunkel,ultra thick] (11.44-0.06,.44+0.06) -- (11.44+0.06,.44-0.06);
	\draw[color=unigrau] (11.75,2.8) node[anchor=south]{\textbf{Demapper}\strut};
    \draw[color=unigrau] (11.75,-1.3) node[anchor=north]{soft-decision metric\strut};
	
	\draw[color=unigrau] (2.125,1.75) node[anchor=south]{\(\mathbf{y}_{\mathrm{CDC},1}\)\strut};
	\draw[color=unigrau] (2.125,-.25) node[anchor=north]{\(\mathbf{y}_{\mathrm{CDC},2}\)\strut};
	\draw[color=unigrau] (4.875,1.75) node[anchor=south]{\(\mathbf{y}_{\mathrm{LP},1}\)\strut};
	\draw[color=unigrau] (4.875,-.25) node[anchor=north]{\(\mathbf{y}_{\mathrm{LP},2}\)\strut};
	\draw[color=unigrau] (7.625,1.25) node[anchor=south]{\(\tilde{\mathbf{s}}_1\)\strut};
	\draw[color=unigrau] (7.625,.25) node[anchor=north]{\(\tilde{\mathbf{s}}_2\)\strut};
	\draw[color=unigrau] (10.375,.75) node[anchor=south]{\(\mathbf{\hat{s}}\)\strut};
	
	\draw[unigrau, fill=uni_gelb!30!white] (8.25,-1.9+.9) rectangle (8.55,-1.6+.9);
	\draw[color=unigrau] (8.43,-1.75+.9) node[anchor=west]{: trainable\strut};
\end{tikzpicture}}
    \caption{Building blocks of transmitter (above) and receiver (below)}
    \label{fig::transceiver}
\end{figure}
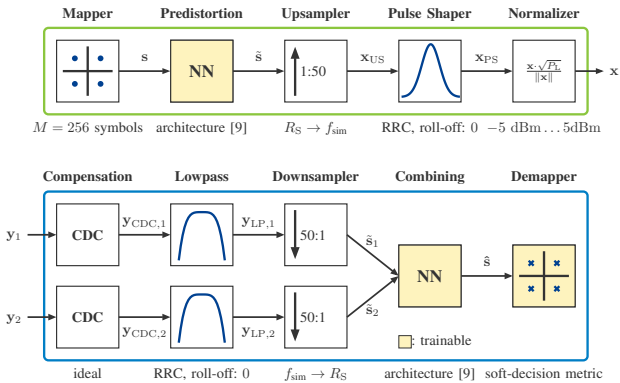

\begin{figure}[t]
    \centering
    \resizebox{.8\linewidth}{!}{\input{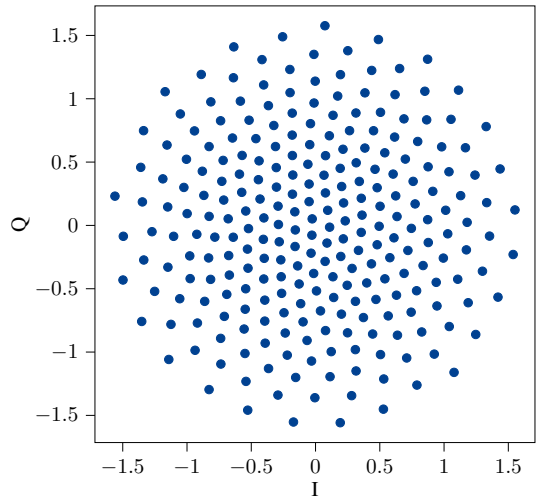}}
    \caption{Used circular constellation}
    \label{fig::constellation}
\end{figure}

Fig. \ref{fig::transceiver} shows the TX (top, green box) and RX (bottom, blue box) used to build the Autoencoder (AE) neural network structure.
Tab.~\ref{tab::system_parameters} provides the simulation parameters used.
\newline
The TX maps the symbol indices onto a geometrically optimized circular constellation as shown in Fig.~\ref{fig::constellation}.
These symbols are then predistorted by a neural network before being upsampled to the simulation frequency.
Pulse shaping is applied and the average power of the signal is normalized.

At the receiver, a chromatic dispersion compensation (CDC) is performed and the signal is downsampled to the symbol rate.
Since the further propagating signal has undergone a different amount of chromatic dispersion, its compensation is performed separately on both paths.
This is indicated by the two parallel paths in the receiver block diagram.
On a symbol-basis, the combined processing of both paths is performed using an NN.
The output is passed on to the trainable demapper, which consists of three fully-connected layers.
Its structure was chosen to give the AE more degrees of freedom during training.
This trainable demapper is used only for training, since a GMM-based and optimized demapper is used for the evaluation.

The architecture for predistortion and combination/compensation is similar to \cite{Freire2021} and adapted to the problem at hand.

The pulse shaper and the low-pass filter before downsampling are assumed to be ideal sinc-functions, i.e., with a brickwall spectrum of bandwidth \(B=20\,\mathrm{GHz}\). %

\begin{table}[b!]
    \centering
    \caption{System parameters}
    \begin{tabular}{R{.3\linewidth}C{.1\linewidth}C{.3\linewidth}}
        \toprule
        \textbf{Property}&\textbf{Symbol}&\textbf{Value}\\
        \midrule
        Attenuation                     &$\alpha$       & \(0.2~\mathrm{dB}\,\mathrm{km}^{-1}\) (\textit{IDRA}) \\
        Chromatic dispersion            &$\beta_2$      & \(-21.67~\mathrm{ps}^2\,\mathrm{km}^{-1}\) \\
        Kerr-effect                     &$\gamma$       & \(1.27~\mathrm{W}^{-1}\,\mathrm{km}^{-1}\) \\
        Carrier frequency               &\(f_0\)        &\(193.55~\mathrm{THz}\)\\
        \multirow{2}{*}{Fiber length}   &\(\ell_1\)     &\(1000~\mathrm{km}\)\\
        &\(\ell_2\)&\textit{to be optimized}\\
        \midrule
        Spontaneous emission factor (Raman)&\(n_{\text{sp},\text{Raman}}\)&\(1.0\)\\
        Spontaneous emission factor (EDFA)&\(n_{\text{sp},\text{EDFA}}\)&\(3.16\)\\
        \midrule
        Simulation frequency            &\(f_\text{sim}\)&\(1~\mathrm{THz}\)\\
        Symbol rate                     &\(R_\mathrm{S}\)&\(20~\mathrm{GBd}\)\\
        Bandwidth DAC/ADC               &\(B\)&\(20~\mathrm{GHz}\)\\
        Launch power                    &\(P_\text{avg}\)&\(-5~\mathrm{dBm}\dots5~\mathrm{dBm}\)\\
        \bottomrule
    \end{tabular}
    \label{tab::system_parameters}
\end{table}

\section{Results\label{sec::results}}
\begin{figure*}
    \centering
    \resizebox{\textwidth}{!}{\begin{tikzpicture}
	\begin{groupplot}[group style={group size=3 by 1,horizontal sep=.2cm}]
		
		\nextgroupplot[
		tick align=outside,
		tick pos=left,
		x grid style={darkgray176},
		xmajorgrids,
		ymajorgrids,
		ymin=0,ymax=0.5,
		width=.45\linewidth,
		height=.33\linewidth,
		title style={yshift=-3pt},
		xtick={0,100,200,300,400,500},
		xticklabels={0,,200,,400,},
		xmin=0,xmax=500,
		xlabel={Length \(\ell_2\) in \(\mathrm{km}\)},
		ylabel={\(\Delta\eta\) in  \(\frac{\mathrm{bit}}{\mathrm{s}\cdot\mathrm{Hz}}\)},
		title={\(P_\mathrm{avg} = 0\,\mathrm{dBm}\)}
		]
		\fill[color=darkgray176,opacity=.5] (axis cs: 320,0.001) rectangle (axis cs: 499,0.499);
		\addplot[very thick,coralplot,mark=o,mark options={solid,line width=.8},mark size=2pt]
		table[col sep=comma, row sep=newline] {fig/se_curves/data/length_plots/gain/aepc-sd_0dBm.txt};
		
		\addplot[very thick,cyanplot,mark=|,mark options={line width=1.2,solid}, mark size=3pt]
		table[col sep=comma, row sep=newline] {fig/se_curves/data/length_plots/gain/aepc-sda_0dBm.txt};

		\nextgroupplot[
		xtick align=outside,
		ytick align=inside,
		yticklabels={},
		ytick style={color=darkgray176},
		tick pos=left,
		x grid style={darkgray176},
		xmajorgrids,
		ymajorgrids,
		ymin=0,ymax=.5,
		xtick={0,100,200,300,400,500},
		xticklabels={0,,200,,400,},
		width=.45\linewidth,
		height=.33\linewidth,
		xmin=0,xmax=500,
		xlabel={Length \(\ell_2\) in \(\mathrm{km}\)},
		title={\(P_\mathrm{avg} = 2.5\,\mathrm{dBm}\)},
		title style={yshift=-3pt},
		legend columns=-1,
		legend style={
			at={(.5,.02)},
			anchor=south,
			draw=black,
			nodes={scale=1, transform shape},
			font=\footnotesize
		},
		legend image post style={xscale=.9}
		]
		\fill[color=darkgray176,opacity=.5] (axis cs: 320,0.001) rectangle (axis cs: 499,0.499);
		\addplot[very thick,coralplot,mark=o,mark options={solid,line width=.8},mark size=2pt] 
		table[col sep=comma, row sep=newline] {fig/se_curves/data/length_plots/gain/aepc-sd_2.5dBm.txt};
		\label{plot::se_gain_sd:sd}
		\addlegendentry{SD}
		\addplot[very thick,cyanplot,mark=|,mark options={line width=1.2,solid}, mark size=3pt] 
		table[col sep=comma, row sep=newline] {fig/se_curves/data/length_plots/gain/aepc-sda_2.5dBm.txt};
		\addlegendentry{SDA}
		\label{plot::se_gain_sd:sda}
		\nextgroupplot[
		xtick align=outside,
		ytick align=inside,
		yticklabels={},
		tick pos=left,
		title style={yshift=-3pt},
		ytick style={color=darkgray176},
		x grid style={darkgray176},
		xmajorgrids,
		ymajorgrids,
		ymin=0,ymax=.5,
		width=.45\linewidth,
		height=.33\linewidth,
		xtick={0,100,200,300,400,500},
		xticklabels={0,,200,,400,},
		xmin=0,xmax=500,
		xlabel={Length \(\ell_2\) in \(\mathrm{km}\)},
		title={\(P_\mathrm{avg} = 5\,\mathrm{dBm}\)},
		]
		\fill[color=darkgray176,opacity=.5] (axis cs: 320,0.001) rectangle (axis cs: 499,0.499);
		
		\addplot[very thick,coralplot,mark=o,mark options={solid,line width=.8},mark size=2pt] 
		table[col sep=comma, row sep=newline] {fig/se_curves/data/length_plots/gain/aepc-sd_5dBm.txt};
		
		\addplot[very thick,cyanplot,mark=|,mark options={line width=1.2,solid}, mark size=3pt] 
		table[col sep=comma, row sep=newline] {fig/se_curves/data/length_plots/gain/aepc-sda_5dBm.txt};

	\end{groupplot}
\end{tikzpicture}}
    \caption{Spectral efficiency gain over length of second fiber \(\ell_2\) (SD: spatial diversity w.o. EDFA, SDA: spatial diversity w. EDFA)}
    \label{fig::length_dependency}
\end{figure*}
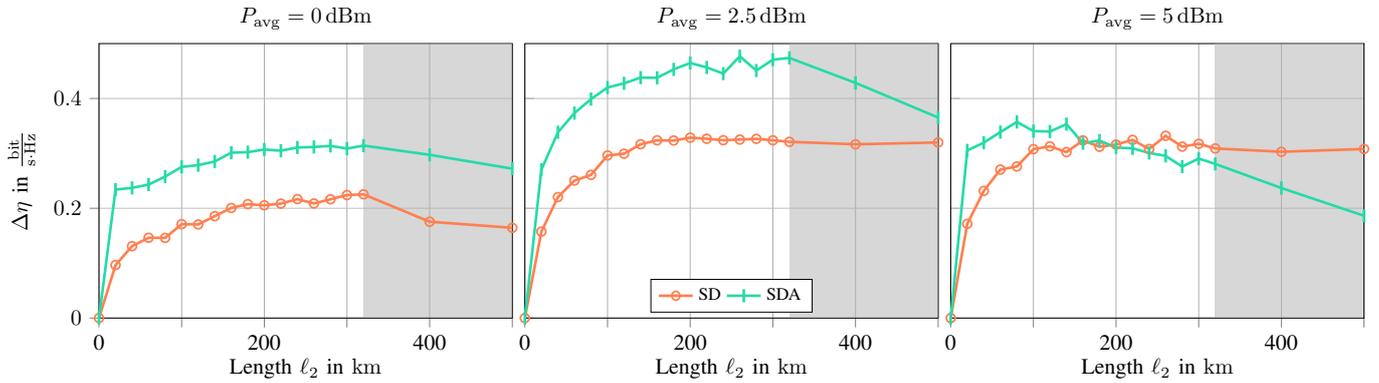
To optimize the two hyperparameters mentioned before, the length of the second fiber \(\ell_2\) and the use of an EDFA to compensate for the coupler loss, we tested our setup with different combinations of both.

For the length we considered a range of up to \(500\,\mathrm{km}\) with a spacing of \(20\,\mathrm{km}\) up to \(320\,\mathrm{km}\) plus the lengths \(400\,\mathrm{km}\) and \(500\,\mathrm{km}\).
The denser grid up to \(320\,\mathrm{km}\) was chosen because a second-order soliton propagating with a power of \(5\,\mathrm{dBm}\) has a periodic length of \(483\,\mathrm{km}\) and previous simulations showed a maximum gain at exactly half that length.
Multiplying this by a factor of \(1.5\) should ensure that the maximum is within this range while \(400\,\mathrm{km}\) and \(500\,\mathrm{km}\) are used as trend indicators for further behavior.

Three different power levels were tested for their ideal length: \(0\,\mathrm{dBm}\), \(2.5\,\mathrm{dBm}\) and \(5\,\mathrm{dBm}\).
Fig.~\ref{fig::length_dependency} shows the gain in terms of spectral efficiency \(\Delta\eta\) compared to a baseline with exactly the same number of parameters in the neural networks.
This baseline does not exploit spatial diversity but obtains the same information for \(\mathbf{y}_1\) and \(\mathbf{y}_2\).
By comparing the systems with spatial diversity to this baseline, we can see the gain achieved purely by exploiting $z$-spatial diversity .
In the plots, SD (\ref{plot::se_gain_sd:sd}) and SDA (\ref{plot::se_gain_sd:sda}) refer to the case with and without splitter loss compensation using an EDFA, respectively.
At first glance, it becomes apparent that SDA outperforms SD at all power levels by up to \(\Delta\eta=+0.15\,\frac{\mathrm{bit}}{\mathrm{s}\cdot\mathrm{Hz}}\).

Both systems, SD and SDA, achieve significant gains of up to \(+0.5\,\frac{\mathrm{bit}}{\mathrm{s}\cdot\mathrm{Hz}}\) over the baseline.
It seems that any small addition of propagation distance in the second fiber produces a large initial gain.
Interestingly, if you compare the curves for SD at \(2.5\,\mathrm{dBm}\) and \(5\,\mathrm{dBm}\), they are almost perfectly aligned and form a sort of plateau after reaching their maximum.
This plateau can also be seen at \(0\,\mathrm{dBm}\), but less pronounced.

Another notable aspect is how the ideal length changes with the input power which is best seen with SDA.
Starting from about \(320\,\mathrm{km}\) at \(0\,\mathrm{dBm}\), the maximum is shifted to a distance of only \(100\,\mathrm{km}\) at \(5\,\mathrm{dBm}\).
This is consistent with soliton theory, where higher power solitons have shorter periodic lengths.

For the \(5\,\mathrm{dBm}\) curve, we observe that the gain achieved with EDFA is partially lower than without EDFA.
There are two possible explanations for this:
Either the maximum gain has already been achieved, indicating that the EDFA is no longer needed at higher input powers.
Thus, the split signal itself still has enough nonlinearity to achieve the same gain. 
Another possible explanation is that the optimization did not converge to a good optimum during training.
This theory can be somewhat discounted since we have seen similar maximum gain curves with spatial diversity in simpler setups, i.e., with fewer parameters and simpler constellations.
In addition, the gain for SD and SDA are quite similar up to \(320\,\mathrm{km}\) supporting the first theory.

In summary, in order to get the maximum performance from exploiting $z$-spatial diversity, the coupler loss should be compensated and the length should be optimized.
Although the performance at a non-ideal length is still superior to the case without EDFA.
For all SDA-curves the gain of the SD case can be achieved with the minimum tested length of \(20\,\mathrm{km}\), which is promising and underlines that most of the gain is already achieved with only a small additional propagation distance with the second fiber.

This allows us to compare spatial diversity systems with conventional systems using CDC and DBP.
We use a split DBP~\cite{Lavery2016} where the predistortion for half of the communication distance \(\ell_1\) is applied at the transmitter and the other half is applied at the receiver.
CDC is applied only at the receiver, as in the AE setup, which therefore should provide a lower performance limit.

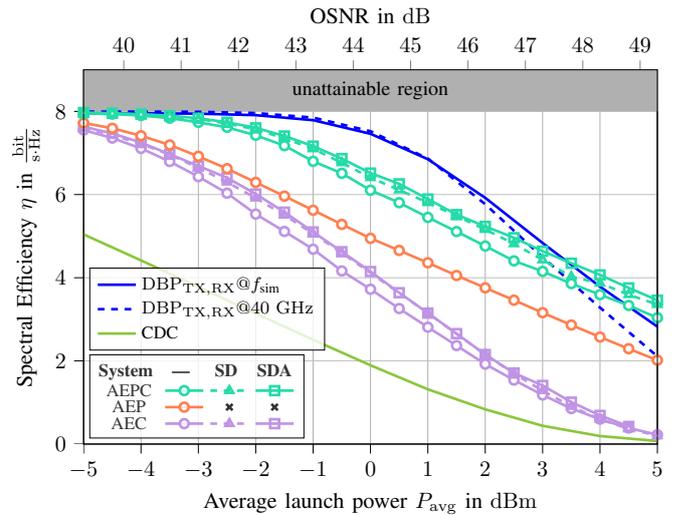
\begin{figure}
    \centering
    \resizebox{\linewidth}{!}{\begin{tikzpicture}
\pgfplotsset{compat=1.3}
	\begin{axis}[
		legend cell align={left},
		legend style={
			fill opacity=0.8,
			draw opacity=1,
			text opacity=1,
			at={(0.01,0.25)},
			anchor=south west,
			draw=black,
			font=\footnotesize
		},
		tick align=outside,
		tick pos=left,
		x grid style={darkgray176},
		xmajorgrids,
		ymajorgrids,
		ymin=0,ymax=9,
		xmin=-5,xmax=5,
		xtick style={color=black},
		ytick style={color=black},
		width=1.2\linewidth,
		height=.85\linewidth,
		xlabel={Average launch power \(P_\mathrm{avg}\) in \(\mathrm{dBm}\)},
		ylabel={Spectral Efficiency \(\eta\) in \(\frac{\mathrm{bit}}{\mathrm{s}\cdot\mathrm{Hz}}\)},
		]
			\fill[color=darkgray176] (axis cs: -30,3.2442) -- (axis cs: -22.5,5.6) -- (axis cs: -15.2,8) -- (axis cs: 10,8) -- (axis cs: 10,16.371) -- (axis cs: -30,16.371) -- (axis cs: -30,3.24419505);

		\addplot[very thick, blue]
		table[col sep=comma, row sep=newline] {fig/se_curves/data/power_plots/dbp_tx_rx_fsim.txt};
		\addlegendentry{\(\mathrm{DBP}_\mathrm{TX,RX}@f_\text{sim}\)}
		
		\addplot[very thick, blue,dashed]
		table[col sep=comma, row sep=newline] {fig/se_curves/data/power_plots/dbp_tx_rx_40GHz.txt};
		\addlegendentry{\(\mathrm{DBP}_\mathrm{TX,RX}@40~\mathrm{GHz}\)}
		
		\addplot[very thick, uni_apfelgruen]
		table[col sep=comma, row sep=newline] {fig/se_curves/data/power_plots/cdc_fsim.txt};
		\addlegendentry{CDC}

		\addplot[very thick, color=lavenderplot,mark=*,mark options={solid,fill=white},forget plot]
		table[col sep=comma, row sep=newline] {fig/se_curves/data/power_plots/aec.txt};
		\label{plot::se_plots_ZOOM:aec}
		
		\addplot[very thick, color=lavenderplot,mark=triangle,mark options={solid,fill=white},forget plot, dashed]
		table[col sep=comma, row sep=newline] {fig/se_curves/data/power_plots/aec-sd.txt};
		\label{plot::se_plots_ZOOM:aec-sd}
		
		\addplot[very thick, color=lavenderplot,mark=square,mark options={solid,fill=white},forget plot]
		table[col sep=comma, row sep=newline] {fig/se_curves/data/power_plots/aec-sda.txt};
		\label{plot::se_plots_ZOOM:aec-sda}

		\addplot[very thick,coralplot,mark=*,mark options={solid,fill=white},forget plot]
		table[col sep=comma, row sep=newline] {fig/se_curves/data/power_plots/aep.txt};
		\label{plot::se_plots_ZOOM:aep}

		\addplot[very thick, mark=x, only marks, unigrau] 
		table[col sep=comma, row sep=newline] {
			10,10
		};
		\label{plot::se_plots_ZOOM:aep_cross}

		\addplot[very thick,cyanplot,mark=*,mark options={solid,fill=white},forget plot]
		table[col sep=comma, row sep=newline] {fig/se_curves/data/power_plots/aepc.txt};
		\label{plot::se_plots_ZOOM:aepc}

		\addplot[very thick,cyanplot,mark=triangle,mark options={solid,fill=white},forget plot,dashed]
		table[col sep=comma, row sep=newline] {fig/se_curves/data/power_plots/aepc-sd.txt};
		\label{plot::se_plots_ZOOM:aepc-sd}

		\addplot[very thick,cyanplot,mark=square,mark options={solid,fill=white},forget plot]
		table[col sep=comma, row sep=newline] {fig/se_curves/data/power_plots/aepc-sda.txt};
		\label{plot::se_plots_ZOOM:aepc-sda}

		\coordinate (legend) at (axis description cs:0.01,.015);
		\draw (axis cs: 0,8.5) node{\small unattainable region};
	\end{axis}
\matrix [
draw,
fill=white,
fill opacity=.8,
matrix of nodes,
align =left,
row sep = -2,
column sep = -2,
inner sep= 2,
anchor=south west,
font=\footnotesize,
mark options={solid}
] at (legend) {
	\textbf{System} & \textbf{---} & \textbf{SD} & \textbf{SDA}\\
	AEPC&\ref{plot::se_plots_ZOOM:aepc}&\ref{plot::se_plots_ZOOM:aepc-sd}&\ref{plot::se_plots_ZOOM:aepc-sda}\\
	AEP&\ref{plot::se_plots_ZOOM:aep}&\ref{plot::se_plots_ZOOM:aep_cross}&\ref{plot::se_plots_ZOOM:aep_cross}\\
	AEC&\ref{plot::se_plots_ZOOM:aec}&\ref{plot::se_plots_ZOOM:aec-sd}&\ref{plot::se_plots_ZOOM:aec-sda}\\
};
	\begin{axis}[
		width=1.2\linewidth,
		height=.85\linewidth,
		tick align=outside,
		ytick pos=right,
		xtick pos=top,
		yticklabel pos=right,
		xticklabel pos=top,
		xlabel={OSNR in \(\mathrm{dB}\)},
		xmin=39.3,xmax=49.3,
		ymin=0,ymax=9,
		ytick={-1},
		xlabel near ticks
		]
	\end{axis}
\end{tikzpicture}}
    \caption{Comparison of spectral efficiency curves versus launch power for different system configurations (AEC: only compensation, AEP: only predistortion, AEPC: predistortion and compensation, SD: spatial diversity w.o. EDFA, SDA: spatial diversity w. EDFA)}
    \label{fig::se_over_power}
\end{figure}

Fig.~\ref{fig::se_over_power} shows the performance in terms of spectral efficiency versus the average launch power for different systems.
We only consider the interval from \(-5\,\mathrm{dBm}\) to \(5\,\mathrm{dBm}\), since the (split) DBP already reaches the maximum achievable performance up to about \(-3\,\mathrm{dBm}\).
The plot shows the performance of three different AE systems: predistortion only at the transmitter (AEP,~\ref{plot::se_plots_ZOOM:aep}), compensation only  at the receiver (AEC,~\ref{plot::se_plots_ZOOM:aec}) and the full system with both predistortion and compensation (AEPC,~\ref{plot::se_plots_ZOOM:aepc}).
Spatial diversity was enabled where applicable, i.e., in systems with compensation.

The comparison of the spatial diversity systems (SD/SDA) with their respective baselines shows that by exploiting the solitonic propagation properties, significant gains can be obtained over the whole power range if we are not limited by saturation.
The gain achieved for AEPC-SDA~(\ref{plot::se_plots_ZOOM:aepc-sda}) compared to the baseline~(\ref{plot::se_plots_ZOOM:aepc}) corresponds to a shift of about \(1\,\mathrm{dBm}\) in terms of average launch power.
Accordingly, the split DBP can already be beaten at \(2.5\,\mathrm{dBm}\) and \(5\,\mathrm{dBm}\) for the DBP operating at \(40\,\mathrm{GHz}\) and \(f_\text{sim}\), respectively.

\section{Summary and conclusion\label{sec::summary}}
In this paper, the use of $z$-spatial diversity for state-of-the-art communication systems has been investigated.
We have shown that significant gains in terms of spectral efficiency (up to almost \(+0.5\,\frac{\mathrm{bit}}{\mathrm{s}\cdot\mathrm{Hz}}\)) can be achieved by combining the two signals.
Simulations show that even higher gains can be achieved by compensating the coupler loss with an EDFA.

Currently, the main issue with this approach is the required length of the additional fiber.
It has already been shown that a length of \(20\,\mathrm{km}\) is sufficient to achieve gains over the baseline.
Fibers designed specifically for this use-case, e.g. highly nonlinear fibers (HNLF) could improve performance for this or even shorter lengths, making the system more practical.
Combining this approach with WDM may also be interesting, since such systems require hard bandwidth limitations and DBP is suboptimal there, leaving more potential for improvement.

\section*{Acknowledgment}
The authors would like to thank D. J. Faul for his preliminary investigation of $z$-spatial diversity with pure solitons.

\end{document}